# An Orbital Solution for WASP-12 b: Updated Ephemeris and Evidence for Decay Leveraging Citizen Science Data

**Avinash S. Nediyedath**
*Department of Physics, Kristu Jayanti College, Bangalore, India, and Exoplanet Watch; avinash123salgunan@gmail.com*

**Martin J. Fowler**
**Anthony Norris**
*Exoplanet Watch*

**Shivaraj R. Maidur**
*Department of Physics, Kristu Jayanti College, Bangalore, India*

**Kyle A. Pearson**
*Jet Propulsion Laboratory, California Institute of Technology, 4800 Oak Grove Drive, Pasadena, CA 91109, and Exoplanet Watch*

**Scott Dixon** **Andre O. Kovacs** **Ken Davis** **Prithwis Das** **Douglas Lalla**
**Pablo Lewin** **Alessandro Odasso** **Michael Primm** **Bryan E. Martin**
*Exoplanet Watch*

**Abstract** NASA Citizen Scientists have used Exoplanet Transit Interpretation Code (EXOTIC) to reduce 40 sets of time-series images of WASP-12 taken by privately owned telescopes and a 6-inch telescope operated by the Center for Astrophysics | Harvard & Smithsonian MicroObservatory (MOBs). Of these sets, 24 result in clean transit light curves of WASP-12 b which are included in the NASA Exoplanet Watch website. We use priors from the NASA Exoplanet Archive to calculate the ephemeris of the planet and combine it with ETD (Exoplanet Transit Database), ExoClock, and TESS (Transiting Exoplanet Survey Satellite) observations. Combining these datasets gives an updated ephemeris for the WASP-12 b system of 2454508.97923 ± 0.000051 BJDTDB with an orbital period of 1.09141935 ± 2.16e–08 days, which can be used to inform the efficient scheduling of future space telescope observations. The orbital decay of the planet was found to be –6.89e–10 ± 4.01e–11 days/epoch. These results show the benefits of long-term observations by amateur astronomers that citizen scientists can analyze to augment the field of exoplanet research.

## 1. Introduction

WASP-12 b was discovered by Hebb *et al.* (2009) and found to be 1.41 times the mass of Jupiter, 1.79 times the radius of Jupiter, and orbiting its F9V host star every 1.09 days. The extreme gravity of the host star is stretching the hot gas giant into an ovoid body, all the while slowly cannibalizing the planet and resulting in a decrease in its orbital period (Yee *et al.* 2020; Wong *et al.* 2022).

The transit method is an important tool for the investigation of Exoplanet systems (Perryman 2018). This method tracks the brightness of the combined system (exoplanet and host star) over time and looks for changes caused when an exoplanet passes in front of its star, which can block some light from reaching the Earth. This technique tells us about the size of the exoplanets and the angle at which they orbit the host star relative to our line of sight. From the observation of multiple transits, it provides information on the orbital period to update the ephemeris. It has become a reliable way of obtaining the mid-transit times of exoplanet orbits. The transit method is within the reach of amateurs with small telescopes, as has been shown by Zellem *et al.* (2020) and Hewitt *et al.* (2023).

In the 14 years since the discovery of WASP-12 b, new tools to investigate exoplanets have been developed, such as the James Webb Space Telescope (JWST). It is being used to study the planets' atmospheric chemistry (Seidel *et al.* 2023). This leads to the need to update the ephemerides of exoplanets to make maximum use of expensive space telescope time to characterize their atmospheres. As of July 2023, we have seen a total of 526 transit observations of WASP-12 b by professional and amateur astronomers in the datasets of ETD (Exoplanet Transit Database, Poddaný *et al.* 2010), ExoClock (Kokori *et al.* 2022), TESS (Ricker *et al.* 2015), and Exoplanet Watch.

In this paper we study 24 transits of WASP-12 b from NASA's Exoplanet Watch, a citizen science project (https://Exoplanets.nasa.gov). Exoplanet Watch enables members around the world to use their time and effort to observe and reduce their own data to produce light curves. We have combined those with observations from the ETD, ExoClock, and TESS databases to update the ephemeris of the exoplanet.

## 2. Observations

Thirty-one observations were made with 60-second, unfiltered exposures with 3-minute cadence collected by a 6-inch aperture MicroObservatory telescope located at Mount Hopkins (latitude 31.675°, longitude –110.952°, 1,268m altitude above sea level) in Arizona. This telescope uses a KAF-1403 ME CCD camera with a pixel scale of 5.2" per pixel and 2 × 2 binning to reduce noise. In addition, nine observations were taken from privately owned telescopes by citizen scientists, yielding a total of 40 observation sets from 03 January 2015 to 06 March 2023. All of the data were analyzed using Exoplanet Transit Interpretation Code (exotic), which is a Python-based tool developed by the Jet Propulsion Laboratory's Exoplanet Watch program for reducing exoplanet transit data. This software can run locally as well as on the cloud via Google's online "Colaboratory" tool (Zellem *et al.* 2020). Prior parameters for WASP-12 b used for nested sampling fitting by exotic are automatically scraped from the NASA Exoplanet Archive (Akeson *et al.* 2013). exotic generates estimates of mid-transit times along with 1σ uncertainties based on the resulting posterior distributions.

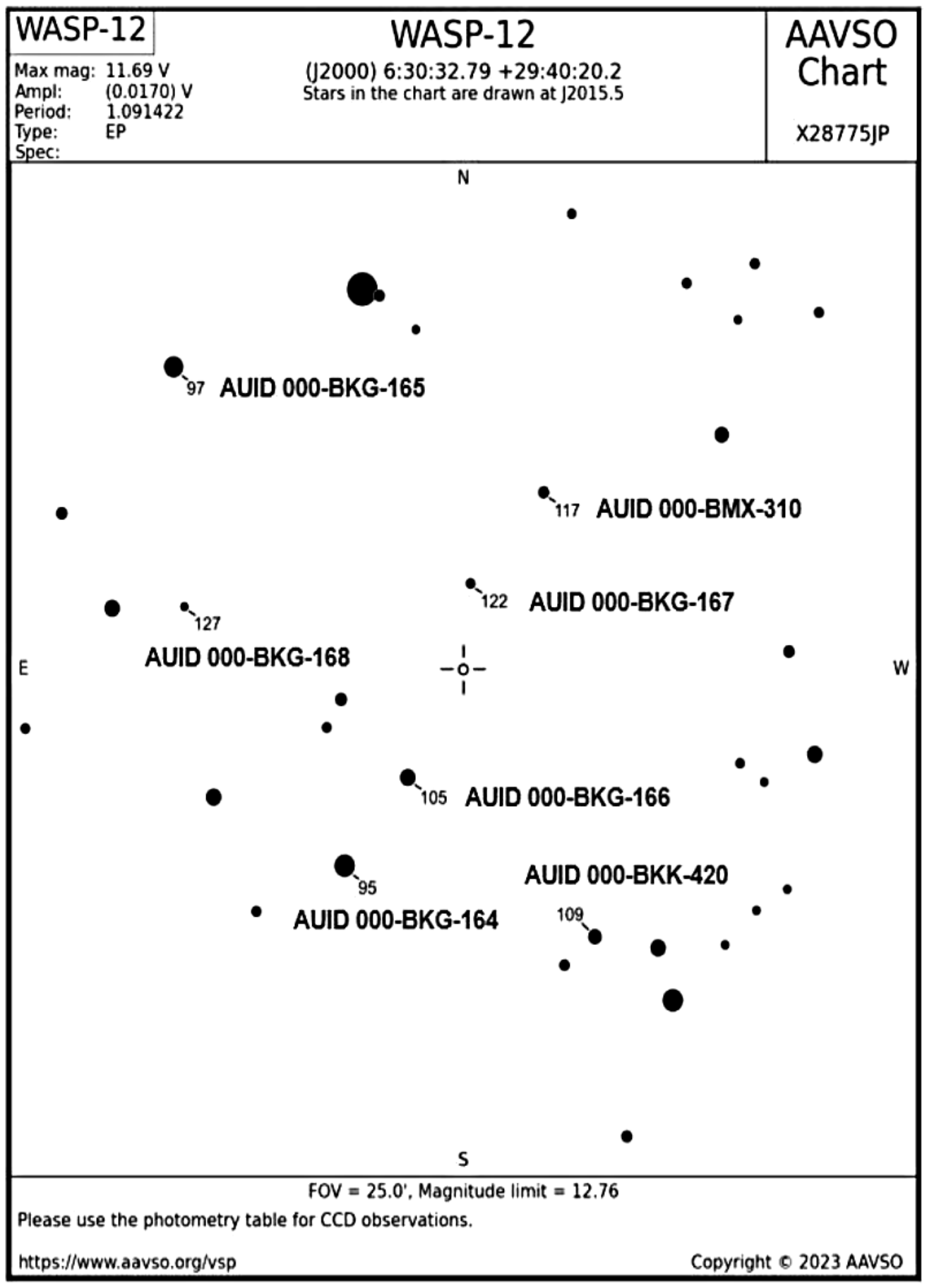

Figure 1. AAVSO VSP view of the WASP-12 star field.

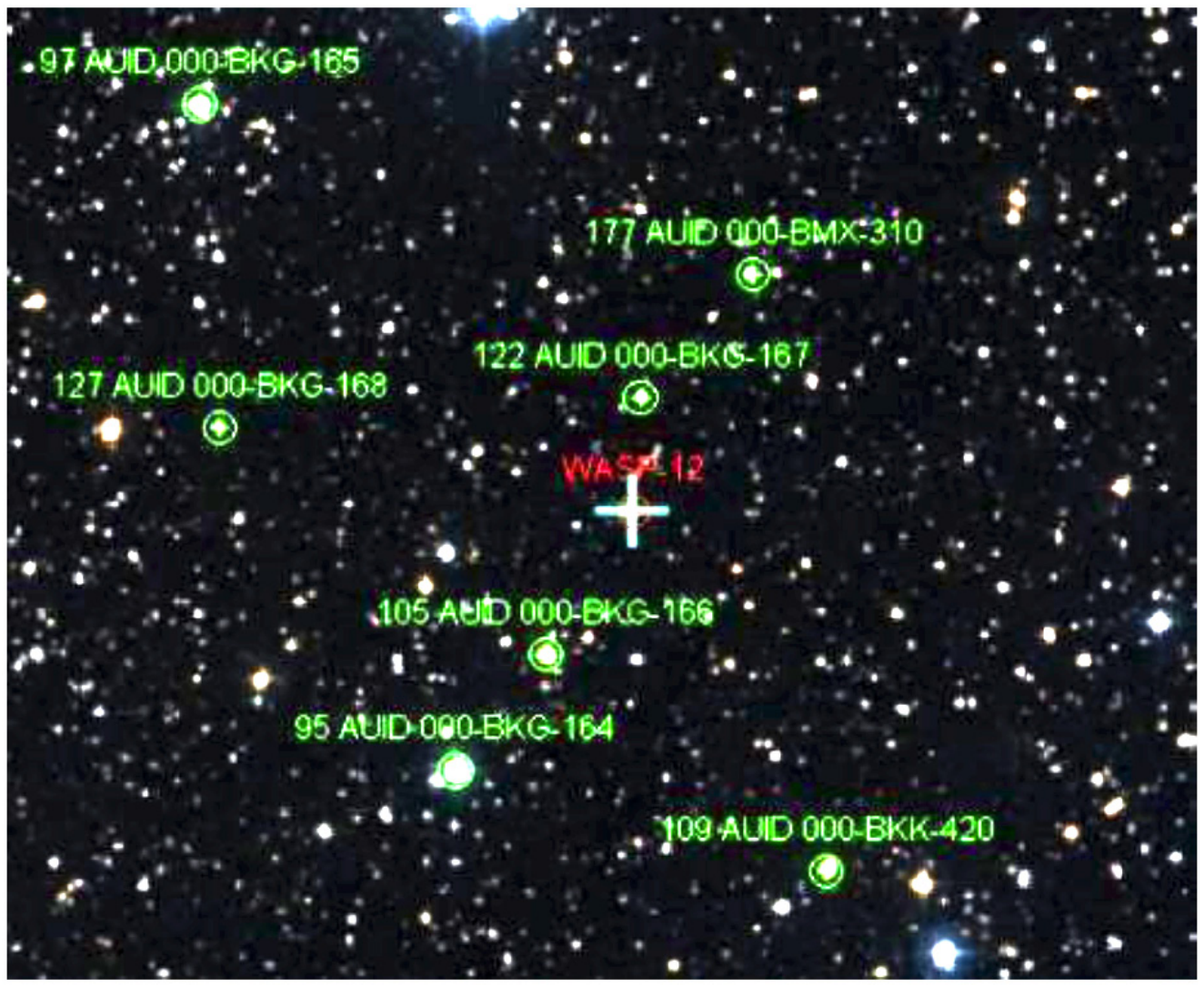

Figure 2. WASP-12 labeled star field in AstroImageJ. Green annotations are used to indicate comparison stars and red annotation is used to indicate the target star. Image by Anthony Norris.

Observations of WASP-12 for reduction were provided by Exoplanet Watch from the MicroObservatory archive for citizen scientists who did not have their own telescope. Using the AAVSO (American Association of Variable Star Observers) finder chart for WASP-12 (see Figure 1), we identified up to seven non-variable comparison stars: AUID 000-BKG-164, AUID 000-BKG-165, AUID 000-BKG-166, AUID 000-BKK-420, AUID 000-BMX-310, AUID 000-BKG-167, and AUID 000-BKG-168. They were selected based on the AAVSO Variable Star Plotter (VSP; AAVSO 2021) and were used for exotic's reduction of the light curves. exotic aligns the images and determines the optimal inner and outer photometric apertures (see Figure 2). The inner aperture encompasses the star's point spread function (PSF) without including the sky background, which fills the space between the outer and inner apertures.

exotic determines the optimal aperture sizes by fitting to a Gaussian PSF model (Mizrachi *et al.* 2021). To account for changes in sky brightness affecting the measured flux, exotic subtracts the background photon count from the star's flux. Finally, the change in flux of the target star is compared to the light emitted by each of the selected comparison stars, and a "quick fit" is performed to identify the best comparison to be used. Nested sampling is used to fit the modeled transit to the observations and produces a triangle plot showing the distribution of posteriors to see whether they were Gaussian (see Figure 3). It is a technique commonly used for posterior exploration and parameter estimation in both ephemeris and light-curve fitting, because of its ability to handle complex parameter spaces and efficiently explore regions of high likelihood. From the sampling, estimates of the full posterior distribution of the parameters are calculated, which is valuable for understanding the uncertainties and correlations between the estimate quantities.

exotic's output included a light curve for each series along with the scatter in the residuals, the midpoint time, transit depth, transit duration, semi-major axis relative to the stellar radius, and planetary versus stellar radius. Example light curves are shown (see Figure 4).

Results from the exotic reductions were uploaded to the AAVSO Exoplanet Database then processed by JPL using the CITISENS (Citizen Initiated Transit Information Survey Enabling NASA Science) pipeline to give the results that are shown on the Exoplanet Watch website and which were used in this study.

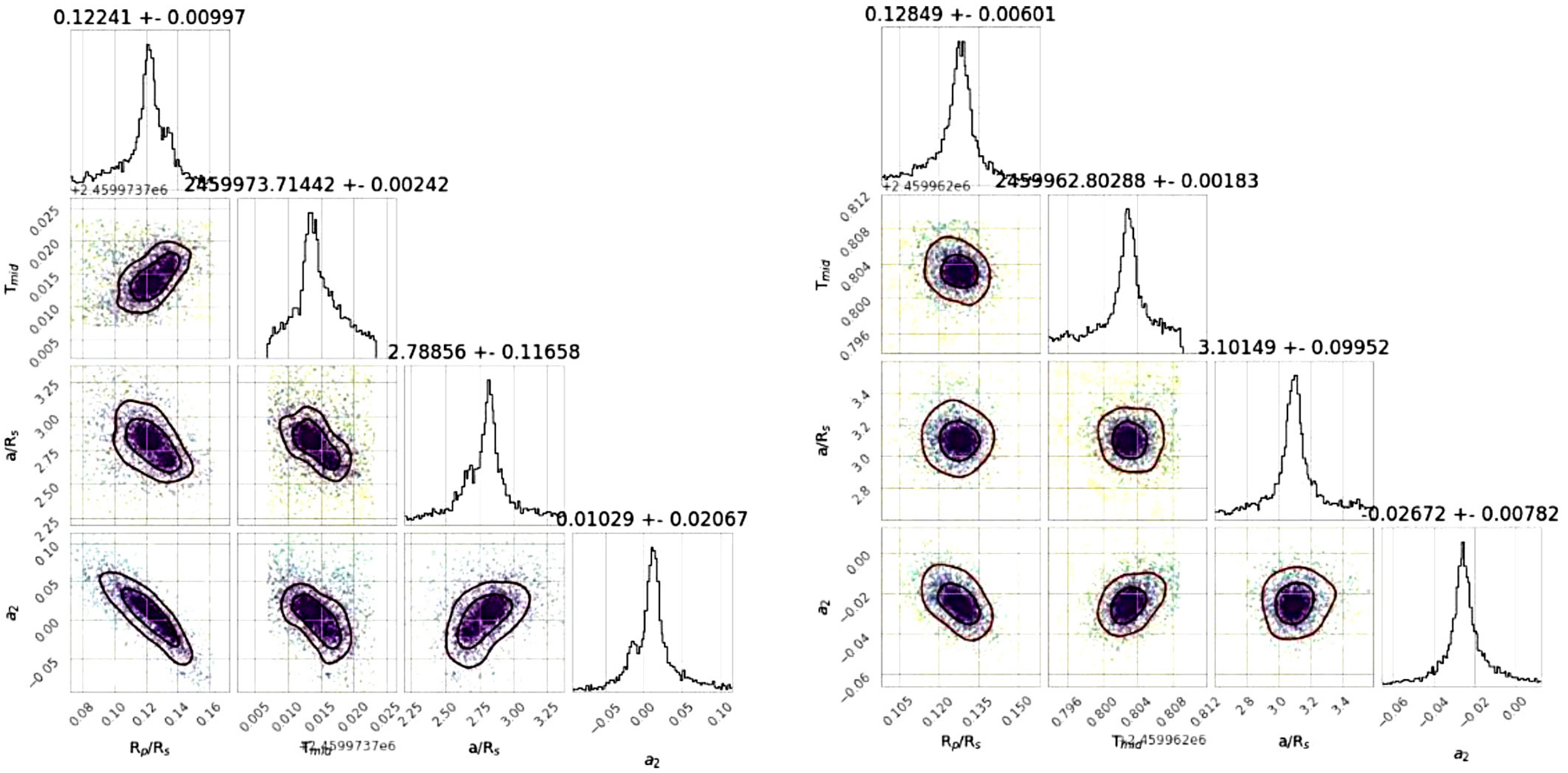

Figure 3. Nested sampling posterior triangle plots using EXOTIC. The data points are color-coded to the likelihood of each fit, with darker colors indicating a higher likelihood. Not all posteriors are shown for reasons of space.

Table 1. Assumed priors by NASA's exoplanet archive for Exoplanet Watch.

| *Parameter* | *Value* | *Uncertainty* | *Units* | *Reference* |
|---|---|---|---|---|
| R.A. 97.624645 | | | Decimal | |
| Dec. 29.6722662 | | | Decimal | |
| Host Star Metallicity | 0.3 | 0.05 | | Öztürk and Erdem (2019) |
| Host Star log(g) | 4.17 | 0.03 | Log10(cgs) | Öztürk and Erdem (2019) |
| Host Star Radius | 1.57 | 0.07 | Sol | Kokori *et al.* (2022) |
| Host Star Effective Temperature | 6300.0 | 200.0 | K | Kokori *et al.* (2022) |
| $a/R_s$ | 3.0 | 0.016 | | Chakrabarty and Sengupta (2019) |
| Eccentricity | 0.0 | 0.01 | | Öztürk and Erdem (2019) |
| Inclination | 83.52 | 0.03 | Deg | Chakrabarty and Sengupta (2019) |
| Omega | 272.7 | 2.4 | Deg | Knutson *et al.* (2014) |
| Orbital Period | 1.09141911 | 6e–08 | Day | Ivshina and Winn (2022) |
| $R_p$ | 21.71 | 0.63 | R_Earth | Chakrabarty and Sengupta (2019) |
| $R_p$ / $R_s$ | 0.1170 | 0.0002 | | Chakrabarty and Sengupta (2019) |
| Ephemeris | 2457010.512173 | 7e–05 | BJDTDB | Ivshina and Winn (2022) |

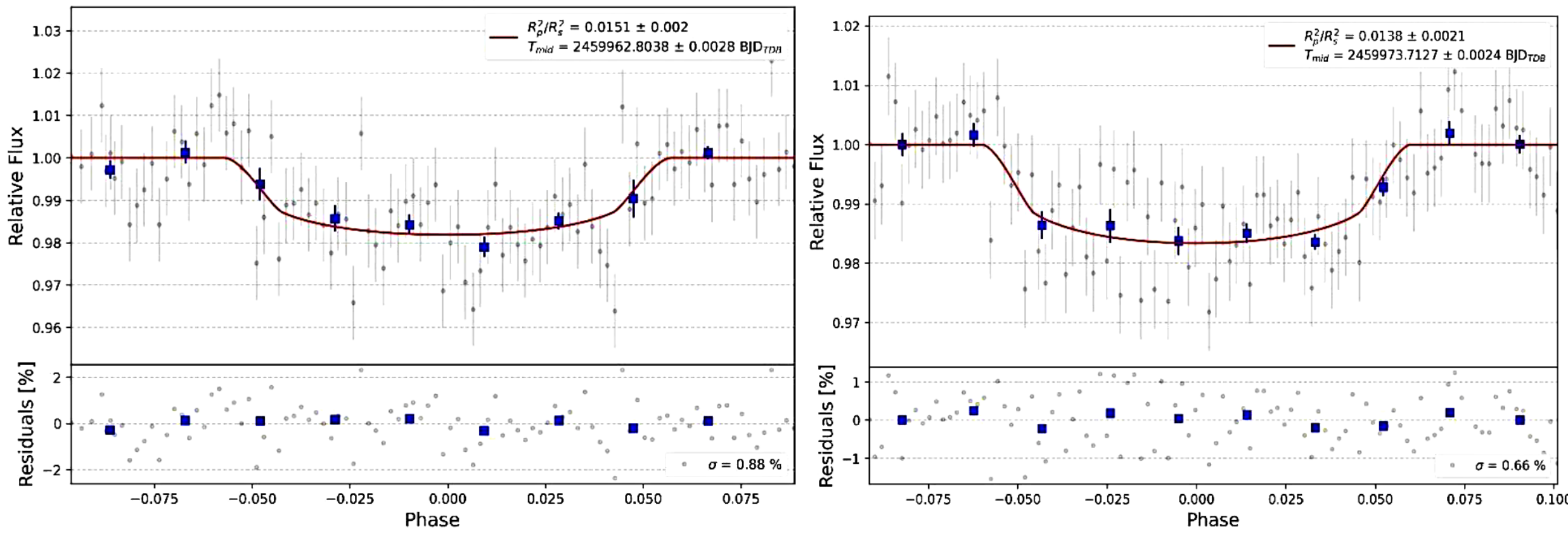

Figure 4. Example transit light curves of WASP-12 b. The gray points represent data from each image in the data set. The blue points represent the average of a set of binned data points, used to guide the eye. The red lines show the EXOTIC model fit for each transit. Not all transits are shown for reasons of space; all the light curves can be seen at: Exoplanet Watch results—Exoplanet Exploration: Planets Beyond our Solar System (nasa.gov).

## 3. Data

There were 14 priors from previously published papers that were used for transit fitting by EXOTIC and CITISENS (see Table 1). EXOTIC's reduction process produced 40 new light curves of WASP-12 b transits (see examples in Figure 4). Of these, seven were duplicate transits taken on the same night by MicroObservatory but which were reduced by different people. There were nine transits that were consistently showing null detection. Adopting the conservative, empirically-derived functions of (Zellem *et al.* 2020) for small telescopes, we consider a transit to have 3-σ detection if $(Rp/Rs)^2$ as a percentage divided by $(R_p / R_s)^2$ uncertainty as a percentage is greater than or equal to 3:

$$(\text{Transit Depth}) / (\text{Transit Depth Uncertainty}) \geq 3 \quad (1)$$

Therefore, a total of 24 observations were taken into account for the Observed-Calculated (O–C) plot (see Table 2). Each point on the plot in Figure 5 shows the observed mid-transit time minus the calculated mid-transit time from the ephemeris along with the combined 1σ uncertainty.

The literature value of 1.657 ± 0.046 solar radii (11.527 × 105 km) for WASP-12 (Chakrabarty and Sengupta 2019) is used for $R_s$ to calculate the radius of the planet in Jupiter radii ($7.149 \times 10^4$ km):

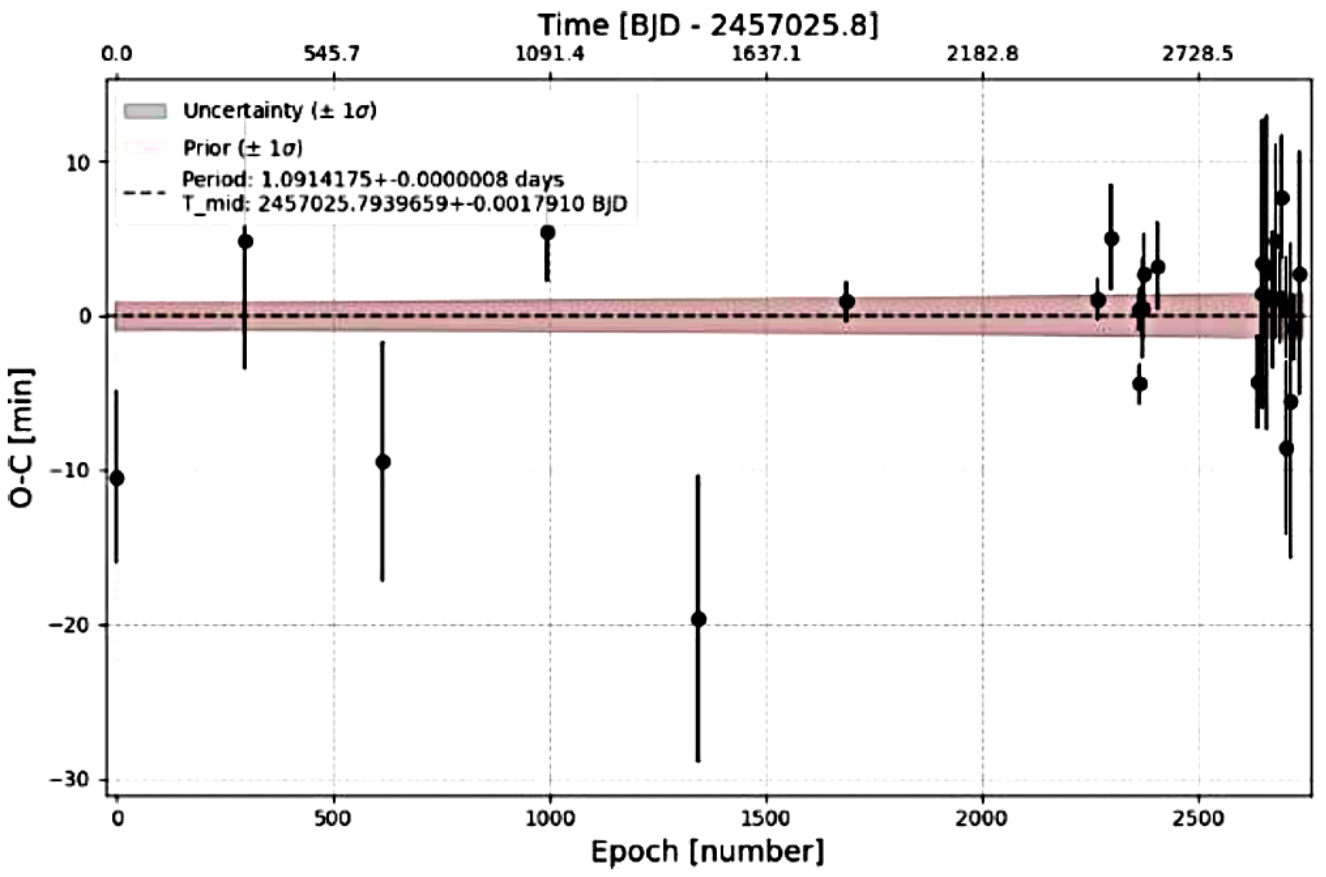

Figure 5. O–C plot of WASP-12 b by Exoplanet Watch.

$$r_j = R_s \times (R_p / R_s) / R_j \quad (2)$$

Here, the planetary size is calculated to be 1.937 ± 0.056 Jupiter radii. Using the MicroObservatory image sets of WASP-12 b transits, we were able to update the ephemeris using the following equation:

$$t_f = n \times P + T_m \quad (3)$$

where $t_f$ is a future mid-transit time, P is the period, n is the orbital epoch, and $T_m$ is a reference mid-transit time. The

Table 2. Exoplanet Watch results for $T_{mid}$ after reduction.

| *Transit Number* | *Date (UTC)* | *Mid-transit (BJDTDB)* | *Mid-transit Uncertainty (days)* | *Observer* | *Observer Code* |
|---|---|---|---|---|---|
| 1 | 2015-01-03 | 2457025.7867 | 0.0038 | Ken Davis | DKEB |
| 2 | 2015-11-21 | 2457347.7655 | 0.0058 | Finn Russom | RFCA |
| 3 | 2016-11-02 | 2457694.8263 | 0.0054 | Prithwis Das | DPRA |
| 4 | 2017-12-25 | 2458112.8501 | 0.0022 | Martin J. Fowler | FMAA |
| 5 | 2019-01-09 | 2458492.6454 | 0.0065 | Martin J. Fowler | FGIC |
| **6** | **2021-11-14** | **2459532.7834** | **0.0024** | **Douglas Lalla** | **LDJC** |
| **7** | **2022-01-25** | **2459604.81036** | **0.00091** | **Mike Chasin** | **CMIA** |
| **8** | **2022-01-25** | **2459604.81379** | **0.00096** | **Douglas Lalla** | **LDJC** |
| **9** | **2022-02-04** | **2459614.6365** | **0.0022** | **Bryan E. Martin** | **MBEB** |
| **10** | **2022-02-06** | **2459616.8209** | **0.0018** | **Scott Dixon** | **DSCC** |
| **11** | **2022-03-13** | **2459651.7466** | **0.002** | **Pablo Lewin** | **LPAC** |
| **12** | **2022-11-20** | **2459903.8588** | **0.0021** | **Anthony Norris** | **NANF** |
| 13 | 2022-12-01 | 2459914.7782 | 0.0066 | Muazzez Kumrucu-Lohmiller | KMUA |
| 14 | 2022-12-02 | 2459915.8686 | 0.0037 | Nathan Kurth | KNAC |
| 15 | 2022-12-02 | 2459925.6921 | 0.0071 | Andre Kovacs | KADB |
| 16 | 2022-12-26 | 2459939.8793 | 0.0031 | Muazzez Kumrucu-Lohmiller | KMUA |
| 17 | 2022-12-26 | 2459949.7043 | 0.0042 | Alessandro Odasso | OAS |
| **18** | **2023-01-14** | **2459959.5249** | **0.0021** | **Andrew Smith** | **SAJB** |
| 19 | 2023-01-18 | 2459962.8037 | 0.0028 | Martin J. Fowler | FMAA |
| 20 | 2023-01-29 | 2459973.7129 | 0.0023 | Martin J. Fowler | FMAA |
| 21 | 2023-01-30 | 2459974.798 | 0.0039 | Martin J. Fowler | FMAA |
| 22 | 2023-02-11 | 2459986.8057 | 0.0071 | Alessandro Odasso | OAS |
| 23 | 2023-02-18 | 2459994.4482 | 0.0014 | Andrew Smith | SAJB |
| **24** | **2023-03-06** | **2460009.731** | **0.0056** | **Michael Primm** | **PMIF** |

*Note: Transits in bold indicate that they did not use MicroObservatory for observations. All transits were used for the O–C plot.*

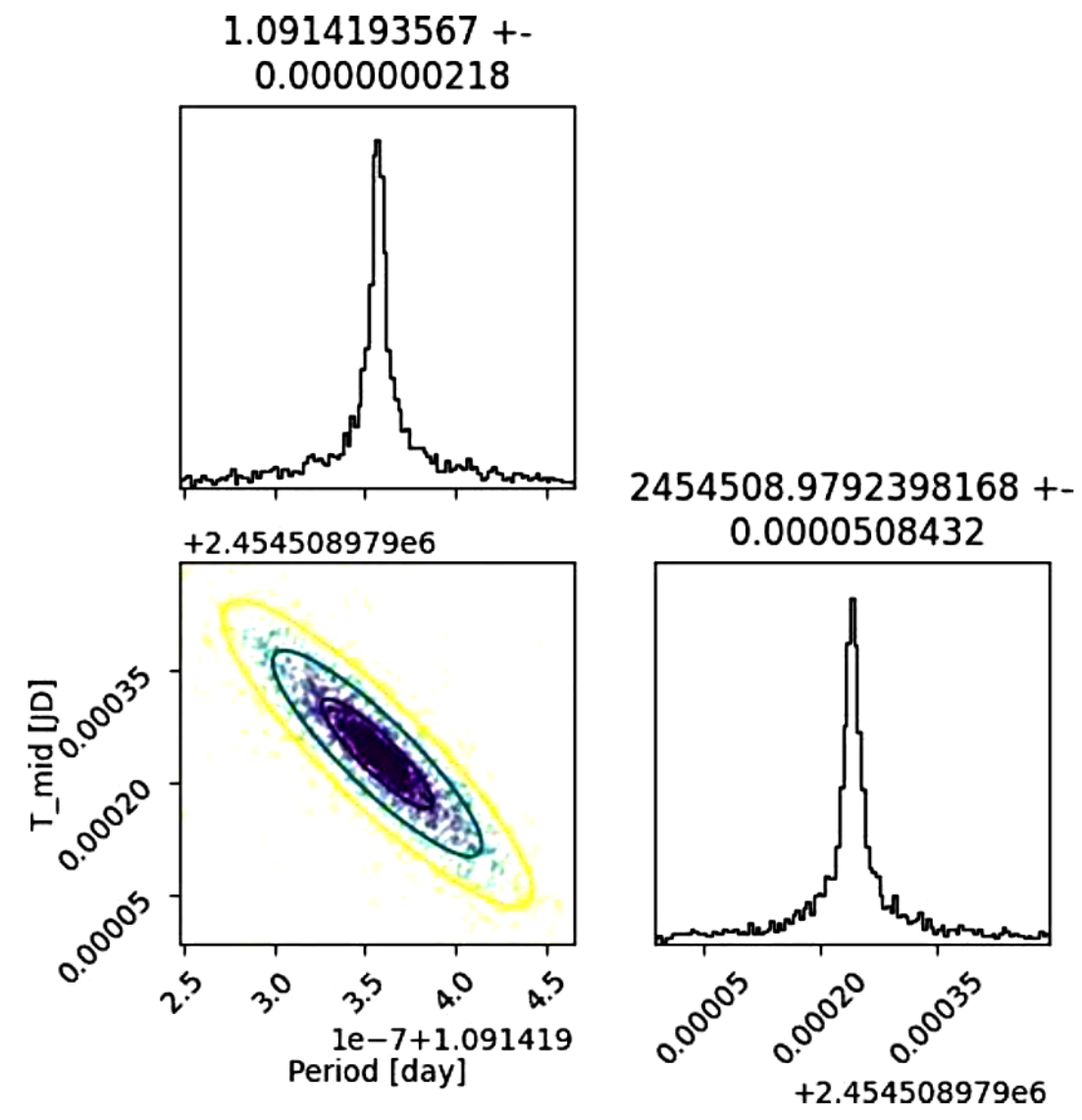

Figure 6. Nested posterior triangle plot using UltraNest for the updated ephemeris. The data points are color- coded to the likelihood of each fit, with darker colors indicating a higher likelihood.

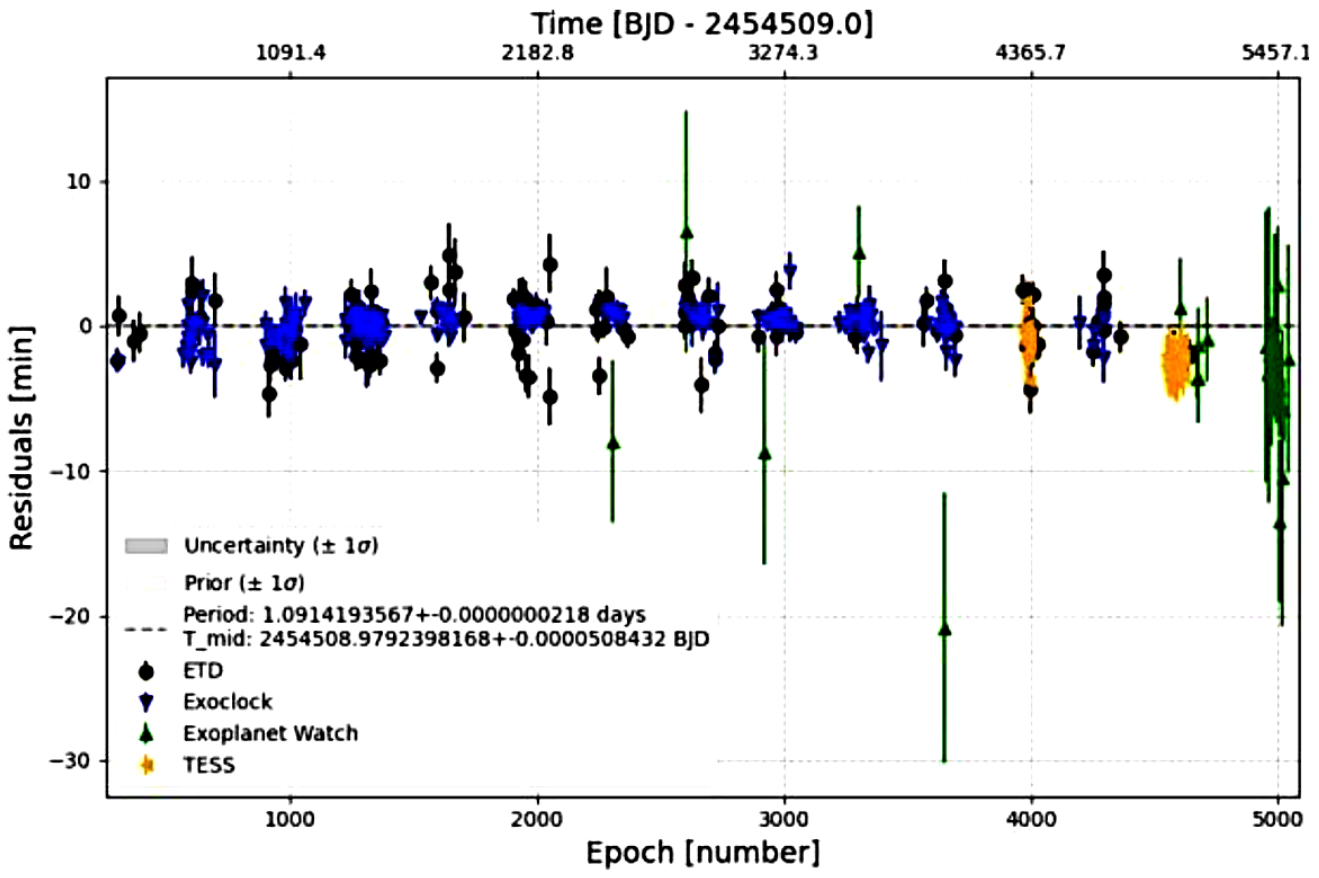

Figure 7. Combined O–C plot data with ExoClock, ETD, TESS, and Exoplanet Watch datasets.

linear ephemeris is optimized using nested sampling to derive posterior distributions for the mid-time and period.

NASA Exoplanet Watch's observations gave a $T_{mid}$ = 2460009.73115 ± 0.00011 BJDTDB with an orbital period of 1.09141889 ± 3.8e–08 days. This is a clear indication of how advanced and easily accessible it has become to reduce transit data from the perspective of a citizen scientist.

Updating the ephemeris of WASP-12 b using amateur observations from Exoplanet Watch can ensure the maximum use is made of expensive non-terrestrial assets such as JWST and ARIEL (Zellem *et al.* 2020, Edwards *et al.* 2019). The ExoClock observations from 12 February 2008 to 20 December 2020 give a $T_{mid}$ = 2457024.706177 ± 5.5e–05 BJDTDB with an orbital period of 1.091419179 ± 4.3e–08 days (Kokori *et al.* 2022). Likewise, ETD observations from 12 February 2008 to 27 December 2021 gave a $T_{mid}$ = 2456594.6766 with an orbital period of 1.09141964 days (Poddaný *et al.* 2010). The ephemerides of the ExoClock and ETD datasets were then forward-propagated using the formula:

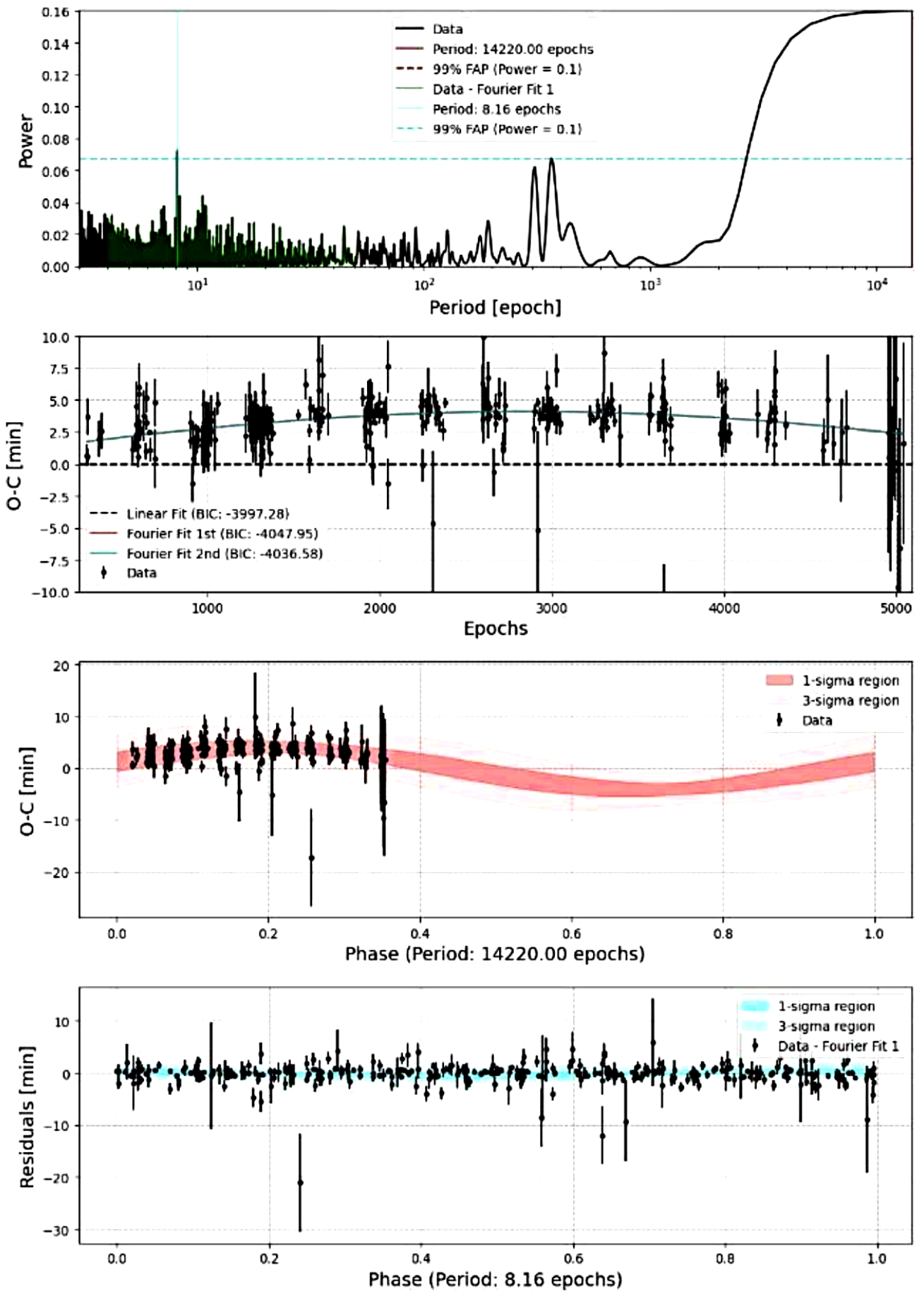

Figure 8. Lomb-Scargle periodogram fitting on the ephemeris to look for period signals.

$$\Delta T_f = (n^2 \cdot \Delta P^2 + 2n \cdot \Delta P \Delta T_m + m^2)^{1/2} \qquad (4)$$

where $T_f$ is a future mid-transit time, n is the orbital epoch, P is the period, and $T_m$ is a reference mid-transit time (Zellem *et al.* 2020).

The 26 TESS observations from 26 December 2019 to 1 December 2021 were then added to the ephemeris. This was done to match the same epoch as Exoplanet Watch to combine the updated ephemerides. Posteriors were then derived for the updated ephemeris of the combined data using nested sampling (see Figure 6) (Pearson *et al.* 2022).

## 4. Results

Combining the Exoplanet Watch, ETD, ExoClock, and TESS datasets gives an updated ephemeris for the WASP-12 b system of 2454508.97923 ± 0.000051 BJDTDB with an orbital period of 1.09141935 ± 2.16e–08 days. This is 0.619-minute different from the original ExoClock dataset, implying that there is a twofold improvement in the precision of the period (see Figure 7). It is clear that the Exoplanet Watch O–C differs from those of ExoClock and ETD in that it appears to have a linear, rather than a non-linear, spread of data points. This is possibly because of the shorter time frame that is covered by the majority of the Exoplanet Watch observations. They extend back only around 500 epochs, compared with the ExoClock and ETD data which cover a longer period of observations (i.e., over 4000 epochs). Whilst this is a linear ephemeris that does not take into account the observed changes over the past 5000 epochs, it may nevertheless be considered sufficiently accurate to inform the efficient scheduling of future space telescope observations.

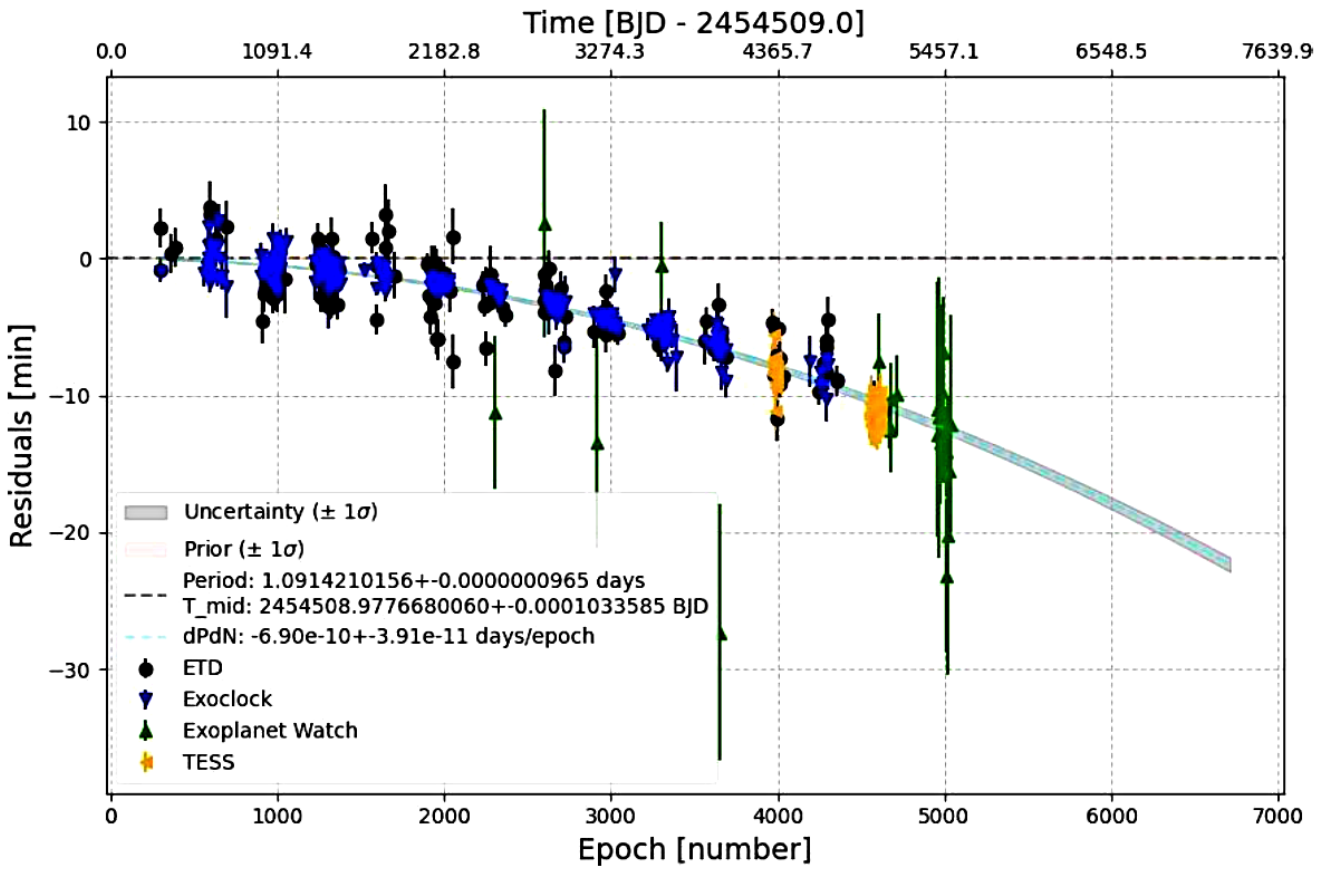

Figure 9. Orbital decay model over the timing residuals of WASP-12 b with future projections around 1σ confidence interval.

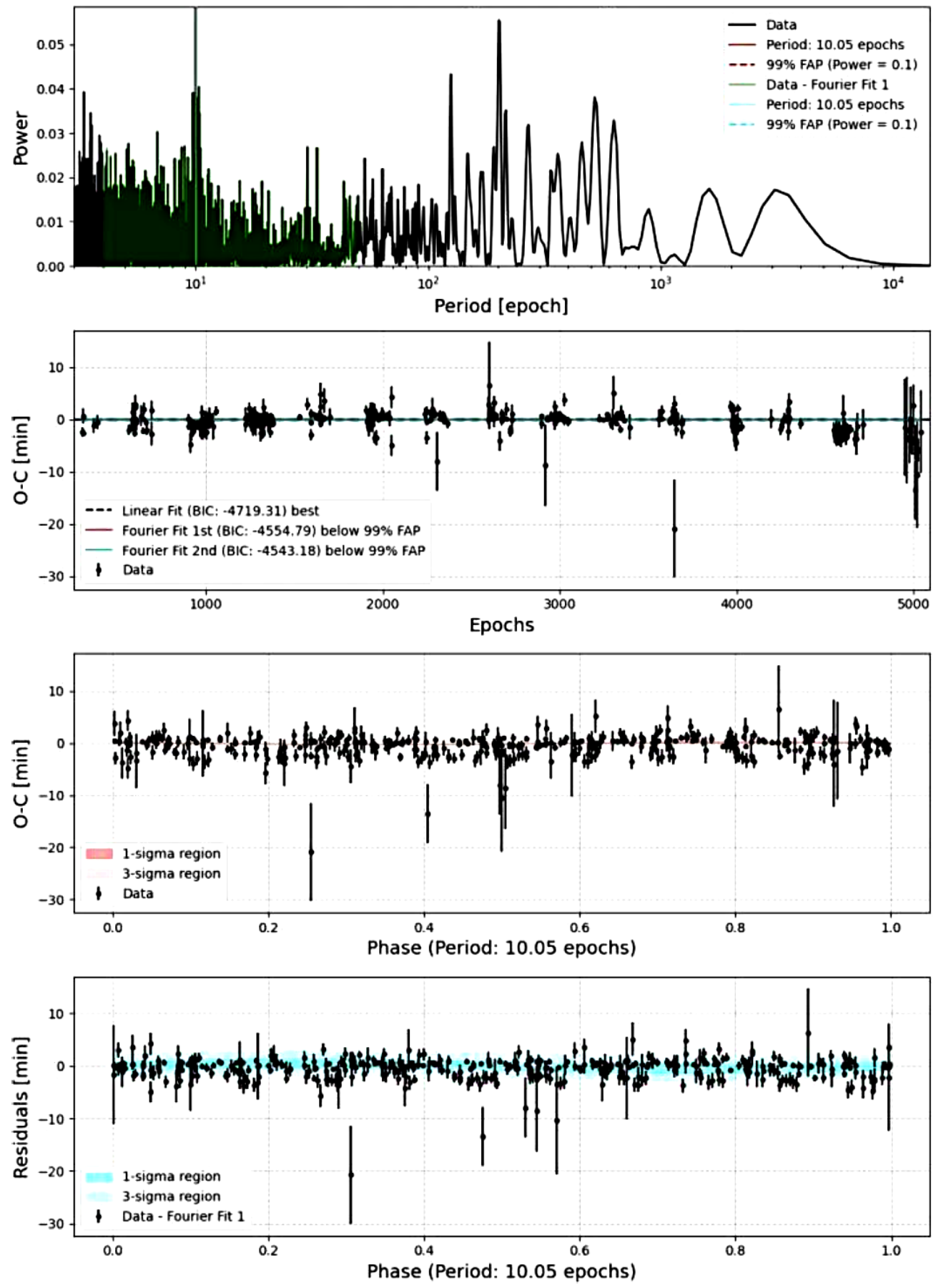

Figure 10. Lomb-Scargle periodogram fitting on the orbital decay model to look for period signals.

This ephemeris was then analyzed with the Lomb-Scargle periodogram to search for periodic signals since the data were irregular with uniform sampling intervals (see Figure 8). Lomb-Scargle is commonly used in astronomy to find periodic signals in light curves generated by variable stars and exoplanets.

The periodogram shows that there is a known system of orbital change which suggests an orbital decay, and which is in agreement with Yee *et al.* (2020), Hagey *et al.* (2022), and Wong *et al.* (2022). The sinusoids are fitted as a linear system of equations to the residuals of the O–C plot. After identifying the signal, it was possible to predict the planet's future behavior after phase-folding it and extending it to twice the period along 1σ and 3σ confidence intervals to a certain extent. It is not possible to completely predict the future behavior since this is not the same as modeling the orbital decay of the planet. In order to overcome this, we modeled the orbital decay of the exoplanet using the equation:

$$t_f = n \times P + T_m + ½ \times d_P / d_n \times n2 \qquad (5)$$

where $t_f$ is a future mid-transit time, P is the period, n is the orbital epoch, and $T_m$ is a reference mid-transit time (Yee *et al.* 2020).

The orbital decay of WASP-12 b was found to be –6.89e–10 ± 4.01e–11 days/epoch. After modeling the orbital decay, we added that model to the linear ephemeris model which can be seen in the new O–C plot (see Figure 9). Using this model it was possible to clearly see the orbital decay and predict the planet's future behavior. Using the new orbital decay model, the Lomb-Scargle periodogram was run again to see if any other periodic signals existed (see Figure 10). No such signals above the 99% FAP (false alarm probability) were found, which indicates that the orbital decay could be the only factor contributing to the change in the orbital period from the previous periodogram.

## 5. Conclusions

This paper presents 24 new mid-transit values and light curves from citizen scientists of Exoplanet Watch for WASP-12 b using MicroObservatory and individual observations. This confirmed parameters for the planet's size and orbit, supporting its classification as a hot Jupiter-type exoplanet. It demonstrates the functionality of EXOTIC and CITISENS and accessibility of its advanced capabilities for use by citizen scientists. We combined Exoplanet Watch, ETD (Poddaný *et al.* 2010), ExoClock (Kokori *et al.* 2022) and TESS datasets to give an updated ephemeris for the WASP-12 b system of 2454508.97923 ± 0.000051 BJDTDB with an orbital period of 1.09141935 ± 2.16e–08 days, which can be used to inform the efficient scheduling of future terrestrial and non-terrestrial observations. The orbital decay of WASP-12 b was found to be –6.89e-10 ± 4.01e-11 days/epoch, which is consistent with other estimates (Yee *et al.* 2020; Wong *et al.* 2022). Further observations can be used to refine this technique and may be used to more precisely determine causes of variations of exoplanet orbits.

## 6. Acknowledgements

Data used here come from the MicroObservatory telescope archives maintained by Frank Sienkiewicz, who also provides information on weather and delta temperature measurements.

MicroObservatory is maintained and operated as an educational service by the Center for Astrophysics | Harvard & Smithsonian and is a project of NASA's Universe of Learning, supported by NASA Award NNX16AC65A. Additional MicroObservatory sponsors include the National Science Foundation, NASA, the Arthur Vining Davis Foundations, Harvard University, and the Smithsonian Institution.

This research has made use of the NASA Exoplanet Archive, which is operated by the California Institute of Technology, under contract with the National Aeronautics and Space Administration under the Exoplanet Exploration Program. This publication makes use of the EXOTIC data reduction package from Exoplanet Watch, a citizen science project managed by NASA's Jet Propulsion Laboratory on behalf of NASA's Universe of Learning. This work is supported by NASA under award number NNX16AC65A to the Space Telescope Science Institute.

This paper includes data collected with the TESS mission, obtained from the MAST data archive at the Space Telescope Science Institute (STScI). Funding for the TESS mission is provided by the NASA Explorer Program. STScI is operated by the Association of Universities for Research in Astronomy, Inc., under NASA contract NAS5–26555.